# Discussions on Inverse Kinematics based on Levenberg-Marquardt Method and Model-Free Adaptive (Predictive) Control

Feilong Zhang

***Abstract*—In this brief, the current robust numerical solution to the inverse kinematics based on Levenberg-Marquardt (LM) method is reanalyzed through control theory instead of numerical method. Compared to current works, the stability of computation and convergence performance of computational error are analyzed much more clearly by analyzing the control performance of the corrected model-free adaptive control (MFAC). Then mainly motivated by minimizing the predictive tracking error, this study suggests a new method of model-free adaptive predictive control (MFAPC) to solve the inverse kinematic problem. At last, we applied the MFAPC as a controller for the robotic kinematic problem. It not only shows an excellent control performance but also efficiently acquires the solution to inverse kinematics.**

## I. INTRODUCTION

There have been tremendous works to solve the inverse kinematics problem [1]-[5]. Among of them, the numerical inverse kinematic solution based on LM [6], [7] method in [1] showed the superior robustness and convergent performance of the computation. It has been utilized in the Robot Toolbox of MATLAB to address the issue of kinematic singularity problem and shows an application prospect in robot controller design. However, this method was just analyzed in numerical method, and did not clearly show the quantitative relationship between the convergence performance of computational error and the damping factor ($\lambda$ in this brief). To this end, we extend the restricted assumptions of the current compact form MFAC [8], [9], and then interestingly find that the corrected MFAC is exactly the numerical method in [1]. Consequently, it is possible for us to analyze the stability and convergence of computation more clearly by the analysis of the system control performance, such as closed-loop function and convergence of tracking error, rather than by the analysis of numerical method in [1].

Then motivated by minimizing the predictive tracking errors over several future steps and its underlying advantages, we apply the MFAPC in the inverse kinematics problem of robot and analyze the system performance by closed-loop system equation. One main merit in comparison to MFAC is that MFAPC can utilize more desired trajectories in the future time, which helps the robot avoid being concentrated with the isolated singular points in generated path. Furthermore, the outputs of the system and controller will be smoother.

Another motivation of this brief is to point out the crucial problem on MFAC in [8]-[12]. They have showed that the tracking error of the system controlled by MFAC converges to zero on the condition that $\lambda$ , i.e., the damping factor in [1], is large enough. This conclusion contradicts with the analysis results in [1] and this brief. For more deficiencies of current works, please refer to [13]-[15].

The rest of the paper is organized as follows. In Section II, the modified compact form equivalent dynamic linearization model (EDLM) is presented for the description of robot kinematics. Based on this, the MFAC controller design and performance analysis are presented. Then we review its application in the Robot Toolbox of MATLAB. Similar to Section III, the MFAPC design and its performance analysis are presented in Section IV, and then how to address inverse kinematic problem by MFAPC is shown. In Section V, we applied MFAPC as a controller for the robotic kinematic control in experiments. It not only shows an excellent control performance but also efficiently acquires the solution to inverse kinematics.

## II. EQUVALENT DYNAMIC LINEARIZATION MODEL FOR MULTIVARIABLE NONLINEAR SYSTEMS

In this section, the robotic kinematic is described by the corrected compact form equivalent dynamic linearization model (EDLM) which is used for controller design and performance analysis.

Consider the kinematic function of robotic manipulator as

$$\boldsymbol{y}(k+1)=\boldsymbol{F}(\boldsymbol{q}(k)) \tag{1}$$

where $\boldsymbol{q}(k)$ represents the system input vector (joint angle vector), and $\boldsymbol{y}(k)$ represents the system output vector (the position and orientation of robot in task space). The dimension of $\boldsymbol{y}(k)$ and $\boldsymbol{q}(k)$ are $M_y$ and $M_u$, respectively. $\boldsymbol{F}(\bullet)$ represents the kinematic transformation from the joint angle vector $\boldsymbol{q}(k)$ to the frame description vector $\boldsymbol{y}(k)$ in Cartesian-space.

According to [14], we make a local approximate linearization of (1) to have the compact form EDLM, as shown by

$$\Delta\boldsymbol{y}(k+1)=\boldsymbol{J}(k)\Delta\boldsymbol{q}(k) \tag{2}$$

Manuscript received Dec 3, 2020. This work was supported in part by the xxxxxxxxxxxxx.

Feilong Zhang is with the State Key Laboratory of Robotics, Shenyang Institute of Automation, Chinese Academy of Sciences, Shenyang 110016, China (e-mail: zhangfeiong@sia.cn).

where $\boldsymbol{J}(k)=\begin{bmatrix} \phi_{11}(k) & \phi_{12}(k) & \cdots & \phi_{1M_u}(k) \\ \phi_{21}(k) & \phi_{22}(k) & \cdots & \phi_{2M_u}(k) \\ \vdots & \vdots & \vdots & \vdots \\ \phi_{M_y1}(k) & \phi_{M_y2}(k) & \cdots & \phi_{M_yM_u}(k) \end{bmatrix} \in \boldsymbol{R}^{My\times Mu}$ ,

$\Delta = 1 - z^{-1}$ in which $z^{-1}$ is the backward shift operator.

## III. MFAC DESIGN AND STABILITY ANALYSIS

This section gives the design and stability analysis methods for MFAC. Then its application in robot inverse kinematic is reviewed.

### *A. Design of Model-Free Adaptive Control*

We rewrite (2) into (3).

$$\boldsymbol{y}(k+1)=\boldsymbol{y}(k)+\boldsymbol{J}(k)\Delta\boldsymbol{q}(k) \tag{3}$$

A control input criterion function is given as:

$$\begin{aligned} J &= \left[\boldsymbol{y}^*(k+1)-\boldsymbol{y}(k+1)\right]^T\left[\boldsymbol{y}^*(k+1)-\boldsymbol{y}(k+1)\right] \\ &\quad +\Delta\boldsymbol{q}^T(k)\boldsymbol{\lambda}\Delta\boldsymbol{q}(k) \end{aligned} \tag{4}$$

where $\boldsymbol{\lambda}=dig(\lambda_1,\cdots,\lambda_{Mu})$ is the weighted diagonal matrix (damping factor matrix in [1]), and we suppose $\lambda_i$ ( $i=1,\cdots,M_u$ ) are equal to $\lambda$ in accordance with [1] and [9]; $\boldsymbol{y}^*(k+1)=\left[y_1^*(k+1),\cdots,y_{My}^*(k+1)\right]^T$ is the desired trajectory vector.

Substituting (3) into (4) and solving the equation $\partial J/\partial\Delta\boldsymbol{q}(k)=\boldsymbol{0}$ yield the MFAC controller:

$$[\boldsymbol{J}^T(k)\boldsymbol{J}(k)+\boldsymbol{\lambda}]\Delta\boldsymbol{q}(k)=\boldsymbol{J}^T(k)[(\boldsymbol{y}^*(k+1)-\boldsymbol{y}(k))] \tag{5}$$

*Remark 1:* If $\boldsymbol{\lambda}=\boldsymbol{0}$, (5) will be the optimal solution for the tracking error control. It was also shown in our previous work [13] for SISO systems.

### *B. Stability Analysis of MFAC*

This section provides the performance analysis of MFAC.

According to (2) and (5), the closed-loop system equation can be expressed by

$$\begin{aligned} &[\Delta\boldsymbol{I}_{My}+z^{-1}\boldsymbol{J}(k)[\boldsymbol{J}^T(k)\boldsymbol{J}(k)+\boldsymbol{\lambda}]^{-1}\boldsymbol{J}^T(k)]\boldsymbol{y}(k) \\ &=\boldsymbol{J}(k)[\boldsymbol{J}^T(k)\boldsymbol{J}(k)+\boldsymbol{\lambda}]^{-1}\boldsymbol{J}^T(k)\boldsymbol{y}^*(k) \end{aligned} \tag{6}$$

where $\boldsymbol{I}_{\bullet}$ represents $\bullet$-dimensional identity matrix.

We assume $\text{rank}\left[\boldsymbol{J}(z^{-1})\right]=M_y$ ( $M_u\geq M_y$ ). By tuning $\boldsymbol{\lambda}$, we can obtain inequation such that

$$\boldsymbol{T}=\Delta\boldsymbol{I}_{My}+z^{-1}\boldsymbol{J}(k)[\boldsymbol{J}^T(k)\boldsymbol{J}(k)+\boldsymbol{\lambda}]^{-1}\boldsymbol{J}^T(k)\neq\boldsymbol{0},\quad |z|>1 \tag{7}$$

, which determines the poles of the system. Then (5) guarantees the system stability according to [18]-[20].

Herein, the singular value decomposition is conducted to $\boldsymbol{J}(k)$ as follow:

$$\boldsymbol{J}(k)=\boldsymbol{U\Sigma V}^T \tag{8}$$

where $\boldsymbol{\Sigma}=[\boldsymbol{\Sigma}_1 \quad \boldsymbol{0}]_{My\times Mu}$, and $\boldsymbol{\Sigma}_1=diag(\sigma_1,\cdots,\sigma_{My})$ is a matrix in which the singular values are diagonally assigned $\sigma_1\geq\sigma_2\geq\cdots\geq\sigma_{My}$; $\boldsymbol{U}$ and $\boldsymbol{V}$ are orthonormal matrices.

Moreover, from (7), we have the poles of the system:

$$\begin{aligned} \boldsymbol{z} &= \boldsymbol{I}_{My}-\boldsymbol{J}(k)[\boldsymbol{J}^T(k)\boldsymbol{J}(k)+\boldsymbol{\lambda}]^{-1}\boldsymbol{J}^T(k) \\ &= \boldsymbol{I}_{My}-\boldsymbol{U\Sigma}[\boldsymbol{\Sigma}^T\boldsymbol{\Sigma}+\boldsymbol{\lambda}]^{-1}\boldsymbol{\Sigma}^T\boldsymbol{U}^T \\ &= \boldsymbol{I}_{My}-\boldsymbol{U\Lambda}(\boldsymbol{U\Lambda})^T \\ &= \boldsymbol{U}(\boldsymbol{I}_{My}-\boldsymbol{\Lambda\Lambda}^T)\boldsymbol{U}^T \end{aligned} \tag{9}$$

where $\boldsymbol{\Lambda}=dig(\frac{\sigma_1}{\sqrt{\lambda+\sigma_1^2}},\cdots,\frac{\sigma_{My}}{\sqrt{\lambda+\sigma_{My}^2}})$.

It is obvious that all the poles of the system will be placed in unit circle under the condition that $\lambda$ is sufficiently large. Moreover, the robustness or stability of the system improves as $\lambda$ increases, including the situation that the robot approaches the singular points. Nevertheless, the convergence of tracking error will be worse, which is demonstrated as follow:

Suppose the desired trajectory is $k^p\bullet\boldsymbol{E}_{My\times1}$ ( $\boldsymbol{E}_{My\times1}=[1,\cdots,1]^T$, $p=1,2,\cdots$ ), the static error (steady-state error) will be

$$\begin{aligned} &\lim_{k\to\infty}\boldsymbol{e}(k) \\ &=\lim_{z\to1}\frac{z-1}{z}(\boldsymbol{I}_{My}-\boldsymbol{T}^{-1}\boldsymbol{J}(k)[\boldsymbol{J}^T(k)\boldsymbol{J}(k)+\boldsymbol{\lambda}]^{-1}\boldsymbol{J}^T(k))\frac{C(z)}{(z-1)^{p+1}}\boldsymbol{E}_{My\times1} \\ &=\lim_{z\to1}\boldsymbol{T}^{-1}(\boldsymbol{I}_{My}-\boldsymbol{J}(k)[\boldsymbol{J}^T(k)\boldsymbol{J}(k)+\boldsymbol{\lambda}]^{-1}\boldsymbol{J}^T(k)\frac{C(z)}{(z-1)^{p-1}}\boldsymbol{E}_{My\times1} \\ &=\lim_{z\to1}\boldsymbol{T}^{-1}\boldsymbol{U}\left[\boldsymbol{I}_{My}-\boldsymbol{\Lambda}^2\right]\boldsymbol{U}^T\frac{C(z)}{(z-1)^{p-1}}\boldsymbol{E}_{My\times1} \\ &=\lim_{z\to1}\boldsymbol{T}^{-1}\boldsymbol{U}dig(\frac{\lambda}{\lambda+\sigma_1^2},\cdots,\frac{\lambda}{\lambda+\sigma_{My}^2})\boldsymbol{U}^T\frac{C(z)}{(z-1)^{p-1}}\boldsymbol{E}_{My\times1} \end{aligned} \tag{10}$$

where $Z(k^p\boldsymbol{E}_{My\times1})=\frac{C(z)}{(z-1)^{p+1}}\boldsymbol{E}_{My\times1}$ and $C(z)$ is the polynomial with the highest power $p$; $Z(\bullet)$ denotes $z$-transformation.

Evidently, the static error and $\boldsymbol{\lambda}$ are positively correlated. Further, through $\boldsymbol{\lambda}=\boldsymbol{0}$ the steady-state error may be eliminated ( $\lim_{k\to\infty}\boldsymbol{e}(k)=\boldsymbol{0}$ ) and the poles of the system will be also placed in coordinate origin to attain the optimal control effect for the tracking error, which means the fastest convergence speed and no overshoot when the robot does not pass through the singular points.

However, the above conclusion contradicts with the result in [8]-[12]. They have showed that the tracking error of the system controlled by MFAC converges to zero on the condition that $\lambda$ is large enough.

### *C. The application of MFAC in robot inverse kinematics*

As to application, someone may refer to the ikine.m of MATLAB Robotic Toolbox. In this part, we will review the theory thereof how to calculate the inverse kinematic solution for the desired position $\boldsymbol{y}^*(k+1)$ in task space. For more details and knowledge, please refer to [16]-[17].

Based on (1), the finite $N$-step forward iteration is given as

$$\boldsymbol{y}(k)_{(i)} = \boldsymbol{F}(\boldsymbol{q}(k)_{(i-1)}), \qquad i = 1, \cdots, N \tag{11}$$

According to [17], we can easily obtain the Jacobian matrix

$$\boldsymbol{J}(k)_{(i)} = \boldsymbol{f}(\boldsymbol{q}(k)_{(i-1)}), \qquad i = 1, \cdots, N-1 \tag{12}$$

where $\boldsymbol{f}(\bullet)$ represents the transformation from the joint angles to the Jacobian matrix. $i$ represents the step of iteration before the robot moves at the time of $k$+1 in experiment; The initializations are $\boldsymbol{y}(k)_{(0)} = \boldsymbol{y}(k)$ and $\boldsymbol{q}(k)_{(0)} = \boldsymbol{q}(k)$. To save the room, the symbol $\boldsymbol{J}(k)_{(i)}$ is shorthand for $\boldsymbol{J}(k+i\,|\,k)$, $\boldsymbol{y}(k)_{(i)}$ for $\boldsymbol{y}(k+i\,|\,k)$ and $\boldsymbol{u}(k)_{(i)}$ for $\boldsymbol{u}(k+i\,|\,k)$, and so on.

Based on (5), we can obtain $\Delta\boldsymbol{q}(k)_{(i)}$ through

$$\begin{aligned}&[\boldsymbol{J}^T(k)_{(i)}\boldsymbol{J}(k)_{(i)} + \boldsymbol{\lambda}(k)_{(i)}]\Delta\boldsymbol{q}(k)_{(i)} \\ &= \boldsymbol{J}^T(k)_{(i)}[(\boldsymbol{y}^*(k+1) - \boldsymbol{y}(k)_{(i)})]\end{aligned} \tag{13}$$

where

$$\boldsymbol{\lambda}(k)_{(i+1)} = \begin{cases} a_1 \bullet \boldsymbol{\lambda}(k)_{(i)}, & \text{if } \dfrac{\left\|\boldsymbol{y}^*(k+1) - \boldsymbol{y}(k)_{(i)}\right\|_2}{\left\|\boldsymbol{y}^*(k+1) - \boldsymbol{y}(k)_{(i-1)}\right\|_2} > 1 \\ \boldsymbol{\lambda}(k)_{(i)} / a_2, & \text{if } \dfrac{\left\|\boldsymbol{y}^*(k+1) - \boldsymbol{y}(k)_{(i)}\right\|_2}{\left\|\boldsymbol{y}^*(k+1) - \boldsymbol{y}(k)_{(i-1)}\right\|_2} \le 1 \end{cases} \tag{14}$$

, in which $a_1 \ge 1$ and $a_2 \ge 1$ are the coefficients for $\boldsymbol{\lambda}(k)_{(i)}$ to make a balance between the convergence speed and the robustness;

Then we have

$$\boldsymbol{q}(k)_{(i)} = \boldsymbol{q}(k)_{(i\text{-}1)} + \Delta\boldsymbol{q}(k)_{(i)} \tag{15}$$

Thus, the iteration among (11)-(15) yields our concerned $\boldsymbol{y}(k)_{(i)}$ and $\boldsymbol{q}(k)_{(i)}$, $i = 1, 2, \cdots, N$ ( $N \le N_{up}$ ). The iteration process above will continue until $N = N_{up}$ or

$$\left\|\boldsymbol{y}^*(k+1) - \boldsymbol{y}(k)_{(N)}\right\|_2 \le \delta \tag{16}$$

where $\delta$ represents the final error tolerance, and $N_{up}$ represents the maximum number of iterations, and the defaults $\delta = 10^{-10}$ and $N_{up} = 500$ are given in ikine.m of MATLAB.

At last, we will send the final iterative inverse kinematic solution (17) to the robot actuator:

$$\boldsymbol{q}(k) = \boldsymbol{q}(k-1) + \sum_{i=1}^{N} \Delta\boldsymbol{q}(k)_{(i)} \tag{17}$$

The presentation of current numerical solution to inverse kinematic is finished.

*Remark 2*: We can modify the criteria (14) to

$$\boldsymbol{\lambda}(k)_{(i+1)} = \begin{cases} a_1 \bullet \boldsymbol{\lambda}(k)_{(i)}, & \text{if } \left\|\boldsymbol{y}^*(k+1) - \boldsymbol{y}(k)_{(i)}\right\|_2 > t_1 \\ \boldsymbol{\lambda}(k)_{(i)} / a_2, & \text{if } \left\|\boldsymbol{y}^*(k+1) - \boldsymbol{y}(k)_{(i)}\right\|_2 \le t_1 \end{cases} \tag{18}$$

where $t_1$ represents the adjustment threshold for $\boldsymbol{\lambda}(k)$.

Further, it is suggested that when $\left\|\boldsymbol{y}^*(k+1) - \boldsymbol{y}(k)_{(r)}\right\|_2$ changing from $\left\|\boldsymbol{y}^*(k+1) - \boldsymbol{y}(k)_{(r)}\right\|_2 \le t_1$ to $\left\|\boldsymbol{y}^*(k+1) - \boldsymbol{y}(k)_{(r+1)}\right\|_2 > t_1$ at step $r$, we may reset $\boldsymbol{\lambda}(k)_{(r)} = \boldsymbol{\lambda}(k)_{(0)}$. And [1] recommends that computation robustly converges with $\boldsymbol{\lambda}(k)_{(0)} = 0.1l^2 \sim 10^{-3}l^2$, and $l$ is the length of a typical link $l = 0.1 \sim 100[m]$.

*Remark 3*: Actually, since the desired position $\boldsymbol{y}^*(k+1)$ is obtained as the known constant vector in the trajectory planning process [16]-[17], we can choose an appropriate constant matrix $\boldsymbol{\lambda}(k)_{(i)} \ne \boldsymbol{0}$ in (13) for easy applications when the robot does not approach the singular points. According to (10), the tracking error still converges when $\boldsymbol{\lambda}(k)_{(i)} = \boldsymbol{0}$, however, the convergence speed is sacrificed inevitably. Alternatively, we can use the lookup table method to tune $\boldsymbol{\lambda}(k)_{(i)}$. When the Jacobian matrix is ill-conditioned or close to singular, we choose a set of large values of $\boldsymbol{\lambda}(k)_{(i)}$ for the computational stability, otherwise, we prefer the small values for the faster convergence speed of the computation. One of our preferred lookup table methods like the Table I is to select the $\boldsymbol{\lambda}(k)_{(i)}$ in accordance with the tracking error which positively correlates with the change from $\boldsymbol{J}(k)_{(i)}$ to $\boldsymbol{J}(k)_{(i+1)}$ and condition number $cond(\boldsymbol{J}(k)_{(i)})$ which normally indicates the degree of ill condition for $\boldsymbol{J}(k)_{(i)}$. The Table I should be determined by trial and error.

TABLE I Matrix $\boldsymbol{\lambda}$ tune table

| $\|\|\boldsymbol{e}\|\|$ \ Cond | $[1, c_1]$ | $[c_1, c_2]$ | $\cdots$ | $[c_{n\text{-}1}, c_n]$ |
|---|---|---|---|---|
| $[0, e_1]$ | $\boldsymbol{0}$ | $\boldsymbol{\lambda}_{12}$ | $\cdots$ | $\boldsymbol{\lambda}_{1n}$ |
| $[e_1, e_2]$ | $\boldsymbol{\lambda}_{21}$ | $\boldsymbol{\lambda}_{22}$ | $\cdots$ | $\boldsymbol{\lambda}_{2n}$ |
| $\vdots$ | $\vdots$ | $\vdots$ | $\cdots$ | $\vdots$ |
| $[e_{m\text{-}1}, e_m]$ | $\boldsymbol{\lambda}_{m1}$ | $\boldsymbol{\lambda}_{m2}$ | $\cdots$ | $\boldsymbol{\lambda}_{mn}$ |

*Remark 4*: Now we want to organize the iteration (11)-(17) into one integrated formula.

Based on (3), the finite $N$-step forward iteration model is given as

$$\begin{aligned} \boldsymbol{y}(k)_{(1)} &= \boldsymbol{y}(k) + \boldsymbol{J}(k)\Delta\boldsymbol{q}(k) \\ \boldsymbol{y}(k)_{(2)} &= \boldsymbol{y}(k)_{(1)} + \boldsymbol{J}(k)_{(1)}\Delta\boldsymbol{q}(k)_{(1)} \\ &= \boldsymbol{y}(k) + \boldsymbol{J}(k)\Delta\boldsymbol{q}(k) + \boldsymbol{J}(k)_{(1)}\Delta\boldsymbol{q}(k)_{(1)} \\ &\vdots \\ \boldsymbol{y}(k)_{(N)} &= \boldsymbol{y}(k)_{(N-1)} + \boldsymbol{J}(k)_{(N-1)}\Delta\boldsymbol{q}(k)_{(N-1)} \\ &= \boldsymbol{y}(k) + \sum_{i=1}^{N}\boldsymbol{J}(k)_{(i-1)}\Delta\boldsymbol{q}(k)_{(i-1)} \end{aligned} \tag{19}$$

which is the linearization of (11). Herein, we define

$$\boldsymbol{\Psi}(k) = \begin{bmatrix} \boldsymbol{J}(k) & & & \\ \boldsymbol{J}(k) & \boldsymbol{J}(k)_{(1)} & & \\ \vdots & \ddots & \ddots & \\ \boldsymbol{J}(k) & \cdots & \boldsymbol{J}(k)_{(N-2)} & \boldsymbol{J}(k)_{(N-1)} \end{bmatrix}_{(N \bullet M_y) \times (N \bullet M_u)},$$

$$\boldsymbol{E}_{(N \bullet My) \times My} = [\boldsymbol{I}_{My} \quad \cdots \quad \boldsymbol{I}_{My}]^T,$$

$$\boldsymbol{Y}_N(k+1) = \begin{bmatrix} \boldsymbol{y}(k)_{(1)} \\ \vdots \\ \boldsymbol{y}(k)_{(N)} \end{bmatrix}_{(N \bullet My) \times 1}, \quad \Delta\boldsymbol{Q}_N(k) = \begin{bmatrix} \Delta\boldsymbol{q}(k) \\ \vdots \\ \Delta\boldsymbol{q}(k)_{(N-1)} \end{bmatrix}_{(N \bullet Mu) \times 1}.$$

Then we can rewrite (19) as

$$\boldsymbol{Y}_N(k+1)=\boldsymbol{E}_{(N\bullet My)\times My}\boldsymbol{y}(k)+\boldsymbol{\Psi}(k)\Delta\boldsymbol{Q}_N(k) \tag{20}$$

Further, we can organize the results calculated by (11)-(16) into one formula

$$\begin{aligned}\Delta\boldsymbol{Q}_N(k)=&[\boldsymbol{\Psi}^T(k)\boldsymbol{\Psi}(k)+\boldsymbol{\lambda}_{(Mu\times N)\times(Mu\times N)}]^{-1}\\ &\bullet\boldsymbol{\Psi}^T(k)[\boldsymbol{Y}^*(k+1)-\boldsymbol{E}_{(N\bullet My)\times My}\boldsymbol{y}(k)]\end{aligned} \tag{21}$$

where $\left[\tilde{\boldsymbol{Y}}^*(k+1)\right]^T=\left[\left(\boldsymbol{y}^*(k+1)\right)^T,\cdots,\left(\boldsymbol{y}^*(k+1)\right)^T\right]$ is the desired system output vector; the diagonal matrix $\boldsymbol{\lambda}_{(Mu\times N)\times(Mu\times N)}=diag(\boldsymbol{\lambda}(k)_{(1)},\cdots,\boldsymbol{\lambda}(k)_{(N)})$ is obtained from (14);

Obviously, when we substitute (20) into (22), (21) will be the optimal solution of (22).

$$\begin{aligned}J=&\left[\boldsymbol{Y}^*(k+1)-\boldsymbol{Y}_N(k+1)\right]^T\left[\boldsymbol{Y}^*(k+1)-\boldsymbol{Y}_N(k+1)\right]\\ &+\Delta\boldsymbol{Q}_N^T(k)\boldsymbol{\lambda}_{(Mu\times N)\times(Mu\times N)}\Delta\boldsymbol{Q}_N(k)\end{aligned} \tag{22}$$

Then (17) is rewritten as

$$\boldsymbol{q}(k)=\boldsymbol{q}(k-1)+\boldsymbol{G}\Delta\boldsymbol{Q}_N(k) \tag{23}$$

where $\boldsymbol{G}_{Mu\times(Mu\bullet N)}=\left[\boldsymbol{I}_{Mu}\quad\cdots\quad\boldsymbol{I}_{Mu}\right]$.

Interestingly, both (11)-(17) or (19) already show the predictive conception. How about we change (13) into the prediction method to minimize the predictive tracking error instead of (4), with naturally utilization of more future desired trajectory (setting points), predictive model and rolling optimization? Motivated by this idea, the MFAPC is studied in Section IV.

## IV. MFAPC DESIGN AND STABILITY ANALYSIS

This section gives the design and stability analysis for MFAPC and its application in inverse kinematic of robot.

### A. *Design of Model-Free Adaptive Predictive Control*

Based on (3), the finite *n*-step forward prediction model is given as

$$\begin{aligned}\boldsymbol{y}(k+1)&=\boldsymbol{y}(k)+\boldsymbol{J}(k)\Delta\boldsymbol{q}(k)\\ \boldsymbol{y}(k+2)&=\boldsymbol{y}(k+1)+\boldsymbol{J}(k+1)\Delta\boldsymbol{q}(k+1)\\ &\quad\boldsymbol{y}(k)+\boldsymbol{J}(k)\Delta\boldsymbol{q}(k)+\boldsymbol{J}(k+1)\Delta\boldsymbol{q}(k+1)\\ &\quad\vdots\\ \boldsymbol{y}(k+n)&=\boldsymbol{y}(k+n-1)+\boldsymbol{J}(k+n-1)\Delta\boldsymbol{q}(k+n-1)\\ &=\boldsymbol{y}(k)+\sum_{i=1}^{n}\boldsymbol{J}(k+i-1)\Delta\boldsymbol{q}(k+i-1)\end{aligned} \tag{24}$$

where *n* is the given prediction step. Then (24) is rewritten as

$$\boldsymbol{Y}_n(k+1)=\boldsymbol{E}_{(n\bullet My)\times My}\boldsymbol{y}(k)+\bar{\boldsymbol{\Psi}}(k)\Delta\boldsymbol{Q}(k) \tag{25}$$

where

$$\bar{\boldsymbol{\Psi}}(k)=\begin{bmatrix}\boldsymbol{J}(k) & & & \\ \boldsymbol{J}(k) & \boldsymbol{J}(k+1) & & \\ \vdots & \ddots & \ddots & \\ \boldsymbol{J}(k) & \cdots & \boldsymbol{J}(k+n-2) & \boldsymbol{J}(k+n-1)\end{bmatrix}_{(n\bullet M_y)\times(n\bullet M_u)},$$

$$\boldsymbol{Y}_n(k+1)=\begin{bmatrix}\boldsymbol{y}(k+1)\\ \vdots\\ \boldsymbol{y}(k+n)\end{bmatrix}_{(n\bullet My)\times 1},\quad \Delta\boldsymbol{Q}(k)=\begin{bmatrix}\Delta\boldsymbol{q}(k)\\ \vdots\\ \Delta\boldsymbol{q}(k+n-1)\end{bmatrix}_{(n\bullet Mu)\times 1},$$

$$\boldsymbol{E}_{(n\bullet My)\times My}=[\boldsymbol{I}_{My}\quad\cdots\quad\boldsymbol{I}_{My}]^T$$

A control input criterion function is given as:

$$\begin{aligned}J=&\left[\boldsymbol{Y}_n^*(k+1)-\boldsymbol{Y}_n(k+1)\right]^T\left[\boldsymbol{Y}_n^*(k+1)-\boldsymbol{Y}_n(k+1)\right]\\ &+\Delta\boldsymbol{Q}^T(k)\bar{\boldsymbol{\lambda}}\Delta\boldsymbol{Q}(k)\end{aligned} \tag{26}$$

where $\bar{\boldsymbol{\lambda}}$ is the weighted diagonal matrix with $\bar{\boldsymbol{\lambda}}=dig(\lambda_1,\cdots,\lambda_{Mu\times n})$ and we suppose all $\lambda_i$ are equal to $\lambda$; $\left[\tilde{\boldsymbol{Y}}_n^*(k+1)\right]^T=\left[\left(\boldsymbol{y}^*(k+1)\right)^T,\cdots,\left(\boldsymbol{y}^*(k+n)\right)^T\right]$ is the desired system output vector and $\boldsymbol{y}^*(k+i)=[y_1^*(k+i),\cdots,y_{My}^*(k+i)]^T$. is the desired system output at the future time $k+i$ ($i$=1,2,⋯, $n$).

We substitute (25) into (26) and solve $\partial J/\partial\Delta\boldsymbol{Q}_n(k)=\boldsymbol{0}$ to have

$$[\bar{\boldsymbol{\Psi}}^T(k)\bar{\boldsymbol{\Psi}}(k)+\bar{\boldsymbol{\lambda}}]\Delta\boldsymbol{Q}(k)=\bar{\boldsymbol{\Psi}}^T(k)[\boldsymbol{Y}_n^*(k+1)-\boldsymbol{E}_{(n\bullet My)\times My}\boldsymbol{y}(k)] \tag{27}$$

Then we retain the current inputs as the local optimal solution:

$$\boldsymbol{q}(k)=\boldsymbol{q}(k-1)+\boldsymbol{g}^T\Delta\boldsymbol{Q}(k) \tag{28}$$

where $\boldsymbol{g}^T=\left[\boldsymbol{I},\boldsymbol{0},\cdots,\boldsymbol{0}\right]$.

### B. *Stability Analysis of MFAC*

This section provides the performance analysis of MFAPC.

From (3), (27) and (28), we have the following closed-loop system equation:

$$\begin{aligned}&[\Delta\boldsymbol{I}_{My}+z^{-1}\boldsymbol{J}(k)\boldsymbol{g}^T[\bar{\boldsymbol{\Psi}}^T(k)\bar{\boldsymbol{\Psi}}(k)+\bar{\boldsymbol{\lambda}}]^{-1}\bar{\boldsymbol{\Psi}}^T(k)\boldsymbol{E}_{(n\bullet My)\times My}]\boldsymbol{y}(k)\\ &=\boldsymbol{J}(k)\boldsymbol{g}^T[\bar{\boldsymbol{\Psi}}^T(k)\bar{\boldsymbol{\Psi}}(k)+\bar{\boldsymbol{\lambda}}]^{-1}\bar{\boldsymbol{\Psi}}^T(k)\boldsymbol{Y}_n^*(k+1)\end{aligned} \tag{29}$$

We assume rank$\left[\boldsymbol{J}(z^{-1})\right]=M_y$ ($M_u\geq M_y$). By tuning $\bar{\boldsymbol{\lambda}}$, we can obtain inequation as follow:

$$\boldsymbol{T}=\Delta\boldsymbol{I}_{My}+z^{-1}\boldsymbol{J}(k)\boldsymbol{g}^T[\bar{\boldsymbol{\Psi}}^T(k)\bar{\boldsymbol{\Psi}}(k)+\bar{\boldsymbol{\lambda}}]^{-1}\bar{\boldsymbol{\Psi}}^T(k)\boldsymbol{E}_{(n\bullet My)\times My}\neq\boldsymbol{0},|z|>1 \tag{30}$$

, which determines the poles of the system. Then (30) guarantees the stability of the system according to [18]-[19].

When the desired trajectory is $\boldsymbol{Y}_n^*(k+1)$ with its component $\boldsymbol{y}^*(k+i)=(k+i-1)^p\bullet[1,\cdots,1]_{1\times My}^T$ ($i$=1,⋯,$n$), the static error will be

$$\begin{aligned}&\lim_{k\to\infty}\boldsymbol{e}(k)\\ &=\lim_{z\to 1}\frac{z-1}{z}(\boldsymbol{I}_{My}-\boldsymbol{T}^{-1}\boldsymbol{J}(k)\boldsymbol{g}^T[\bar{\boldsymbol{\Psi}}^T(k)\bar{\boldsymbol{\Psi}}(k)+\bar{\boldsymbol{\lambda}}]^{-1}\bar{\boldsymbol{\Psi}}^T(k)\boldsymbol{P})\frac{C(z)}{\left(z-1\right)^{p+1}}\\ &=\lim_{z\to 1}\boldsymbol{T}^{-1}(\Delta\boldsymbol{I}_{My}-\boldsymbol{J}(k)\boldsymbol{g}^T[\bar{\boldsymbol{\Psi}}^T(k)\bar{\boldsymbol{\Psi}}(k)+\bar{\boldsymbol{\lambda}}]^{-1}\bar{\boldsymbol{\Psi}}^T(k)\\ &\quad\bullet(\boldsymbol{P}-z^{-1}\boldsymbol{E}_{(n\bullet My)\times My}))\frac{C(z)}{\left(z-1\right)^{p}}\end{aligned} \tag{31}$$

where

$$
\begin{aligned}
z(\boldsymbol{Y}_n^*(k+1)) &= diag(\boldsymbol{I}_{My},\cdots,z^n\boldsymbol{I}_{My})\boldsymbol{E}_{(n\bullet My)\times My}z(\boldsymbol{y}^*(k+1)) \\
&= \boldsymbol{P}\frac{C(z)}{(z-1)^{p+1}}z(\boldsymbol{y}^*(k+1))
\end{aligned} \tag{32}
$$

and $\boldsymbol{P}^T=[\boldsymbol{I}_{My},z\boldsymbol{I}_{My},\cdots,z^{n-1}\boldsymbol{I}_{My}]$. Further, under the condition that the robot does not pass through the singular points, through choosing $\bar{\boldsymbol{\lambda}}=\boldsymbol{0}$ the steady-state error will be eliminated since

$$
\begin{aligned}
\lim_{k\to\infty}\boldsymbol{e}(k) &= \lim_{z\to 1}\boldsymbol{T}^{-1}(\boldsymbol{I}_{My}-\boldsymbol{J}(k)\boldsymbol{g}^T[\bar{\boldsymbol{\Psi}}(k)]^{-R}\boldsymbol{E}_{(n\bullet My)\times My})\frac{C(z)}{(z-1)^{p-1}} \\
&= \lim_{z\to 1}\boldsymbol{T}^{-1}(\boldsymbol{I}_{My}-\boldsymbol{J}(k)\boldsymbol{g}^T[\bar{\boldsymbol{\Psi}}(k)]^{-R}\boldsymbol{E}_{(n\bullet My)\times My})\frac{C(z)}{(z-1)^{p-1}} \\
&= \lim_{z\to 1}\boldsymbol{T}^{-1}(\boldsymbol{I}_{My}-\boldsymbol{J}(k)\left[\boldsymbol{J}(k)]^{-R},\boldsymbol{0},\cdots,\boldsymbol{0}\right]\boldsymbol{E}_{(n\bullet My)\times My})\frac{C(z)}{(z-1)^{p-1}}=\boldsymbol{0}
\end{aligned} \tag{33}
$$

where

$$
[\bar{\boldsymbol{\Psi}}(k)]^{-R}=
\begin{bmatrix}
[\boldsymbol{J}(k)]^{-R} & \boldsymbol{0} & \cdots & & \boldsymbol{0} \\
-[\boldsymbol{J}(k+1)]^{-R} & [\boldsymbol{J}(k+1)]^{-R} & \boldsymbol{0} & & \boldsymbol{0} \\
\boldsymbol{0} & -[\boldsymbol{J}(k+2)]^{-R} & [\boldsymbol{J}(k+2)]^{-R} & \cdots & \boldsymbol{0} \\
\vdots & \vdots & \ddots & \ddots & \vdots \\
\boldsymbol{0} & \cdots & \boldsymbol{0} & -[\boldsymbol{J}(k+n-1)]^{-R} & [\boldsymbol{J}(k+n-1)]^{-R}
\end{bmatrix}
$$

and $[\bullet]^{-R}$ represents the right inverse matrix of $\bullet$.

### C. *The application of MFAPC in robot inverse kinematics*

The application of MFAPC is similar to that of MFAC in Section III.

The finite $n$-step forward prediction model at the $i$-th iteration is given as

$$
\boldsymbol{y}(k+j)_{(i)}=\boldsymbol{F}(\boldsymbol{q}(k+j)_{(i-1)}),\qquad j=0,\cdots,n-1 \tag{34}
$$

where $i=1,\cdots,N$. According to [17], we can easily obtain the Jacobian matrix

$$
\boldsymbol{J}(k+j)_{(i)}=\boldsymbol{f}(\boldsymbol{q}(k+j)_{(i-1)}),\qquad j=0,\cdots,n-1 \tag{35}
$$

Then we let

$$
\bar{\boldsymbol{\Psi}}(k)_{(i)}=
\begin{bmatrix}
\boldsymbol{J}(k)_{(i)} & & & \\
\boldsymbol{J}(k)_{(i)} & \boldsymbol{J}(k+1)_{(i)} & & \\
\vdots & \ddots & \ddots & \\
\boldsymbol{J}(k)_{(i)} & \cdots & \boldsymbol{J}(k+n-2)_{(i)} & \boldsymbol{J}(k+n-1)_{(i)}
\end{bmatrix}
$$

Based on (27), we can obtain $\Delta\boldsymbol{Q}(k)_{(i)}$ through

$$
\begin{aligned}
&[\bar{\boldsymbol{\Psi}}^T(k)_{(i)}\bar{\boldsymbol{\Psi}}(k)_{(i)}+\bar{\boldsymbol{\lambda}}(k)_{(i)}]\Delta\boldsymbol{Q}(k)_{(i)} \\
&=\bar{\boldsymbol{\Psi}}^T(k)_{(i)}[\boldsymbol{Y}_n^*(k+1)-\boldsymbol{E}_{(n\bullet My)\times My}\boldsymbol{y}(k)_{(i)}]
\end{aligned} \tag{36}
$$

And we have tuned $\bar{\boldsymbol{\lambda}}(k)_{(i+1)}$ in accordance with (37) in the experiment.

$$
\bar{\boldsymbol{\lambda}}(k)_{(i)}=
\begin{cases}
\boldsymbol{0} & \text{if} \quad 1<\mathrm{c}(\boldsymbol{J}(k)_{(i)})<c_1 \\
\bar{\boldsymbol{\lambda}}_1 & \text{if} \quad c_1\le \mathrm{c}(\boldsymbol{J}(k)_{(i)})<c_2 \\
\vdots & \\
\bar{\boldsymbol{\lambda}}_n & \text{if} \quad \mathrm{c}(\boldsymbol{J}(k)_{(i)})\ge c_n
\end{cases} \tag{37}
$$

where $\mathrm{c}(\boldsymbol{J}(k)_{(i)})=\max\left\{cond(\boldsymbol{J}(k)_{(i)}),\cdots,cond(\boldsymbol{J}(k+n-1)_{(i)})\right\}$, and $c_i$ ($i$=1,⋯,$n$) are the constants. If we choose $\bar{\boldsymbol{\lambda}}(k)_{(i)}=\boldsymbol{0}$ to reduce the amount of calculation, we will have

$$
[\bar{\boldsymbol{\Psi}}^T(k)_{(i)}\bar{\boldsymbol{\Psi}}(k)_{(i)}+\bar{\boldsymbol{\lambda}}(k)_{(i)}]^{-1}\bar{\boldsymbol{\Psi}}^T(k)_{(i)}=[\bar{\boldsymbol{\Psi}}(k)_{(i)}]^{-R}=
$$

$$
\begin{bmatrix}
[\boldsymbol{J}(k)_{(i)}]^{-R} & \boldsymbol{0} & \cdots & & \boldsymbol{0} \\
-[\boldsymbol{J}(k+1)_{(i)}]^{-R} & [\boldsymbol{J}(k+1)_{(i)}]^{-R} & \boldsymbol{0} & & \boldsymbol{0} \\
\boldsymbol{0} & -[\boldsymbol{J}(k+2)_{(i)}]^{-R} & [\boldsymbol{J}(k+2)_{(i)}]^{-R} & \cdots & \boldsymbol{0} \\
\vdots & \vdots & \ddots & \ddots & \vdots \\
\boldsymbol{0} & \cdots & \boldsymbol{0} & -[\boldsymbol{J}(k+n-1)_{(i)}]^{-R} & [\boldsymbol{J}(k+n-1)_{(i)}]^{-R}
\end{bmatrix} \tag{38}
$$

Then we have

$$
\boldsymbol{q}(k)_{(i)}=\boldsymbol{q}(k)_{(i-1)}+\boldsymbol{g}^T\Delta\boldsymbol{Q}(k)_{(i)} \tag{39}
$$

The iteration described by (34)-(39) yields our concerned $\boldsymbol{Y}_n(k+1)_{(i)}\triangleq[\boldsymbol{y}^T(k)_{(i)},\cdots,\boldsymbol{y}^T(k+n-1)_{(i)}]^T$ and $\boldsymbol{q}(k)_{(i)}$, $i$=1,⋯, $N$ ($N\le N_{up}$). The above iteration will continue until $N=N_{up}$ or

$$
\left\|\boldsymbol{Y}_n^*(k+1)-\boldsymbol{Y}_n(k+1)_{(N)}\right\|_2\le\delta \tag{40}
$$

At last, we will send the final iterative inverse kinematic solution (41) to the robot actuator.

$$
\boldsymbol{q}(k)=\boldsymbol{q}(k-1)+\sum_{i=1}^{N}\Delta\boldsymbol{q}(k)_{(i)} \tag{41}
$$

*Remark 5*: Though (33) demonstrates that the tracking error converges to zero on the condition that $\bar{\boldsymbol{\lambda}}=\boldsymbol{0}$, the motivation and effect of MFAPC is to make a balance between the minimization of $\left\|\boldsymbol{Y}_n^*(k+1)-\boldsymbol{Y}_n(k+1)\right\|_2$ and the system robustness. As to the difference between MFAPC and MFAC, MFAC is to make a balance between the minimization of $\left\|\boldsymbol{y}^*(k+1)-\boldsymbol{y}(k+1)\right\|_2$ and the system robustness. If we let $\boldsymbol{Y}_n^*(k+1)=\boldsymbol{E}\boldsymbol{y}^*(k+1)$, the MFAPC will have the same control effect as the MFAC. On the other hand, through $n=1$ the MFAPC degenerates into the MFAC. Therefore, the MFAPC incorporates the MFAC method in [1] and [8].

## V. SIMULATION

Example 1: In this example, the MFAPC is used as a controller (27), (28) for the robotic kinematic control problem. By this way, we can not only verify its control performance but also exhibit its effectiveness on acquiring the solution to inverse kinematic.

We consider a robot with three links $l_1$=5, $l_2$= $l_3$=7 as shown in Fig. 1.

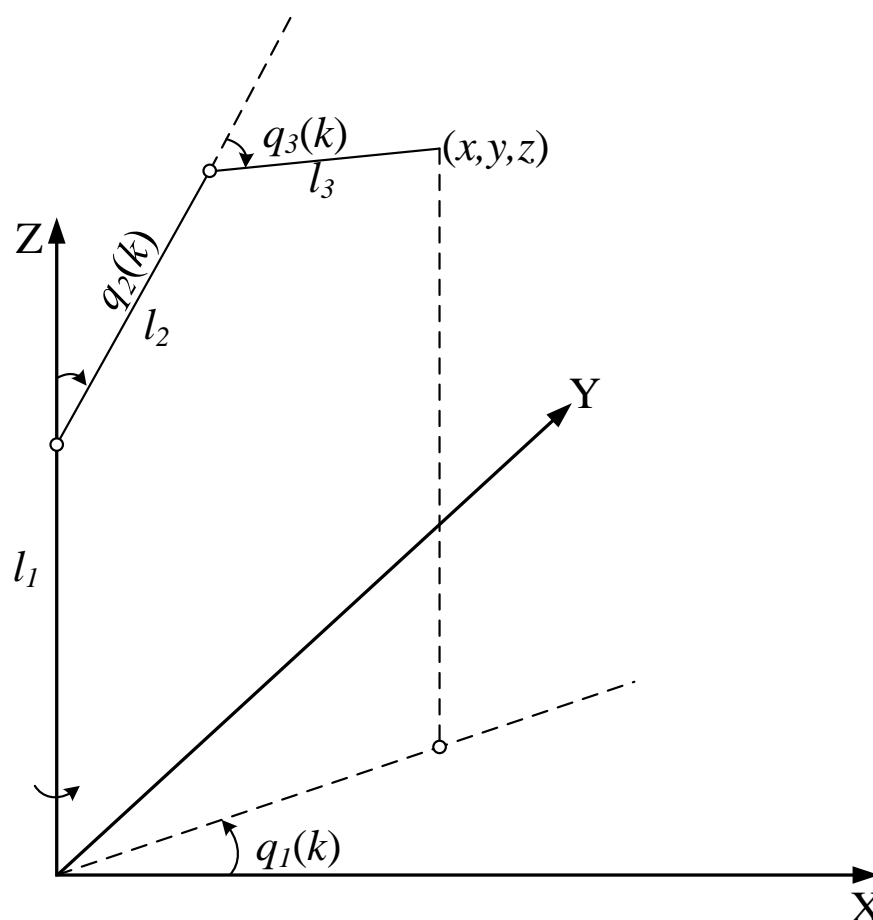

Fig. 1 Robot model

The outputs of the system are the position $(x,y,z)$ of the robot manipulator in task space. The inputs of the system are the angles $q_1(t)$, $q_2(t)$, $q_3(t)$ of the robot in joint space. The system model is

$$
\begin{aligned}
x(k+1) &= (l_2\sin(q_2(k)) + l_3\sin(q_2(k)+q_3(k))\cos(q_1(k)) \\
y(k+1) &= (l_2\sin(q_2(k)) + l_3\sin(q_2(k)+q_3(k))\sin(q_1(k)) \\
z(k+1) &= l_1 + l_2\cos(q_2(k)) + l_3\cos(q_2(k)+q_3(k))
\end{aligned} \tag{42}
$$

We take partial derivation of equation (42) to have the equivalent-dynamic-linearization model:

$$
\Delta \boldsymbol{y}(k+1) = \begin{bmatrix} \Delta x(k+1) \\ \Delta y(k+1) \\ \Delta z(k+1) \end{bmatrix} = \boldsymbol{J}(k) \begin{bmatrix} \Delta q(k) \\ \Delta q_2(k) \\ \Delta q_3(k) \end{bmatrix} \tag{43}
$$

where $\boldsymbol{J}(k)$ represents the (pseudo) Jacobian matrix. The desired output trajectory is considered as a helical curve:

$$
\begin{aligned}
x^*(k) &= 4 + 3\sin(\pi k/50) \\
y^*(k) &= 3\cos(\pi k/50) \qquad 1 \le k \le 800 \\
z^*(k) &= 5 + k/200
\end{aligned}
$$

The initial values are $[x(1), y(1), z(1)] = 0$, $q_1(1) = q_2(1) = q_3(1) = 0$, nevertheless these initial settings do not suffice the actual forward kinematics of the robot system . The initial controller parameter is $\lambda = 2$;

In order to make a balance between the convergence of the tracking error and the system stability, the MFAPC with prediction step $n$=5 is applied as follows:

$$
\begin{aligned}
\Delta \boldsymbol{u}(k) = \boldsymbol{g}^T [\bar{\boldsymbol{\Psi}}^T(k)\bar{\boldsymbol{\Psi}}(k) + \lambda \boldsymbol{I}]^{-1} \bar{\boldsymbol{\Psi}}^T(k) \\
\bullet [\boldsymbol{Y}^*(k+1) - \boldsymbol{E}_{(N \bullet My) \times My} \boldsymbol{y}(k)]
\end{aligned} \tag{44}
$$

$$\text{If } \left\| \boldsymbol{Y}_n^*(k+1) - \boldsymbol{Y}_n(k+1) \right\|_2 > 10,$$

$$\lambda(k) = 1.1 \bullet \lambda(k-1)$$

$$\text{else } \lambda(k) = \lambda(k-1)/1.02$$

where $\boldsymbol{y}^*(k+1) = [x^*(k+1), y^*(k+1), z^*(k+1)]^T$ is the desired trajectory, and the current position is $\boldsymbol{y}(k) = [x(k), y(k), z(k)]^T$ . We make a local linear approximation $\bar{\boldsymbol{\Psi}}(k) = \begin{bmatrix} \boldsymbol{J}(k) & & & \\ \boldsymbol{J}(k) & \boldsymbol{J}(k) & & \\ \vdots & \ddots & \ddots & \\ \boldsymbol{J}(k) & \cdots & \boldsymbol{J}(k) & \boldsymbol{J}(k) \end{bmatrix}_{(n \bullet M_y) \times (n \bullet M_u)}$ in this simulation.

The outputs of the system controlled by MFAC are shown in Fig. 2. The control inputs and the value of controller parameter $\lambda$ are shown in Fig. 3. Fig. 4 shows the elements in $\boldsymbol{J}(k)$ .

Since the initial value of inputs and outputs of the system violates the kinematics of robot, the beginning tracking performance is not well. Simultaneously, the $\lambda$ increases to enhance the robustness of the system. After the system is stable with the tracking error of the system $\left\| \boldsymbol{Y}_n^*(k+1) - \boldsymbol{Y}_n(k+1) \right\|_2$ lower than 10 at time of 33, the $\lambda$ decreases to accelerate the convergence of the tracking error.

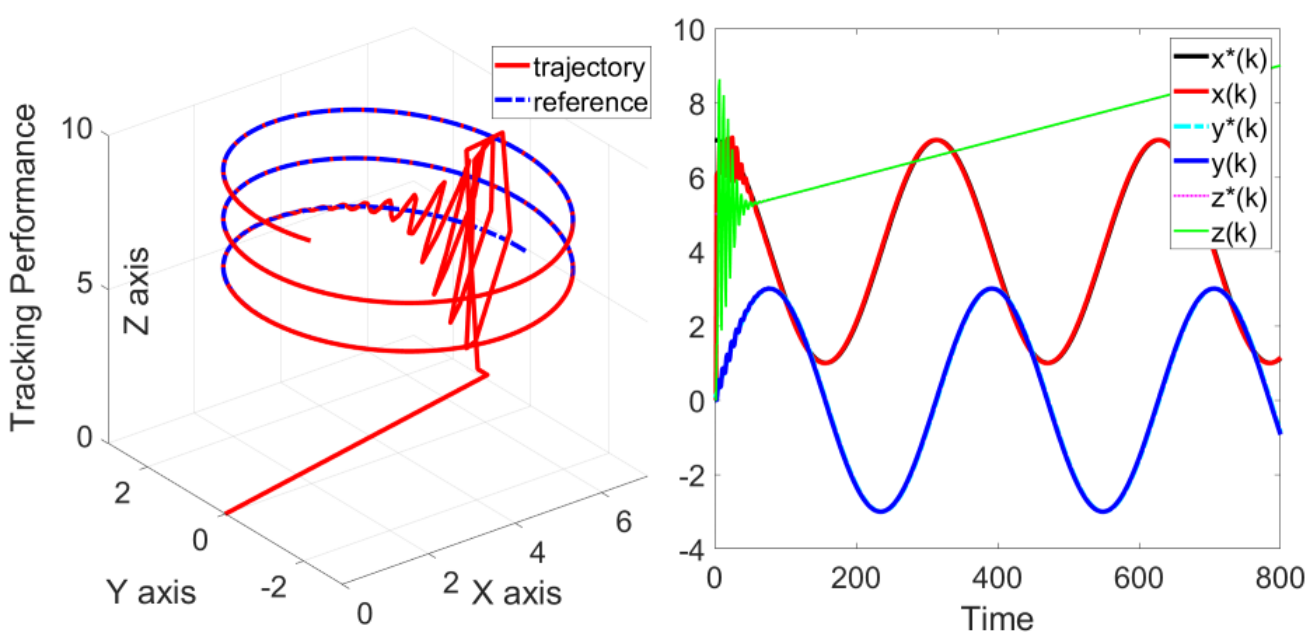

Fig. 2 Tracking performance

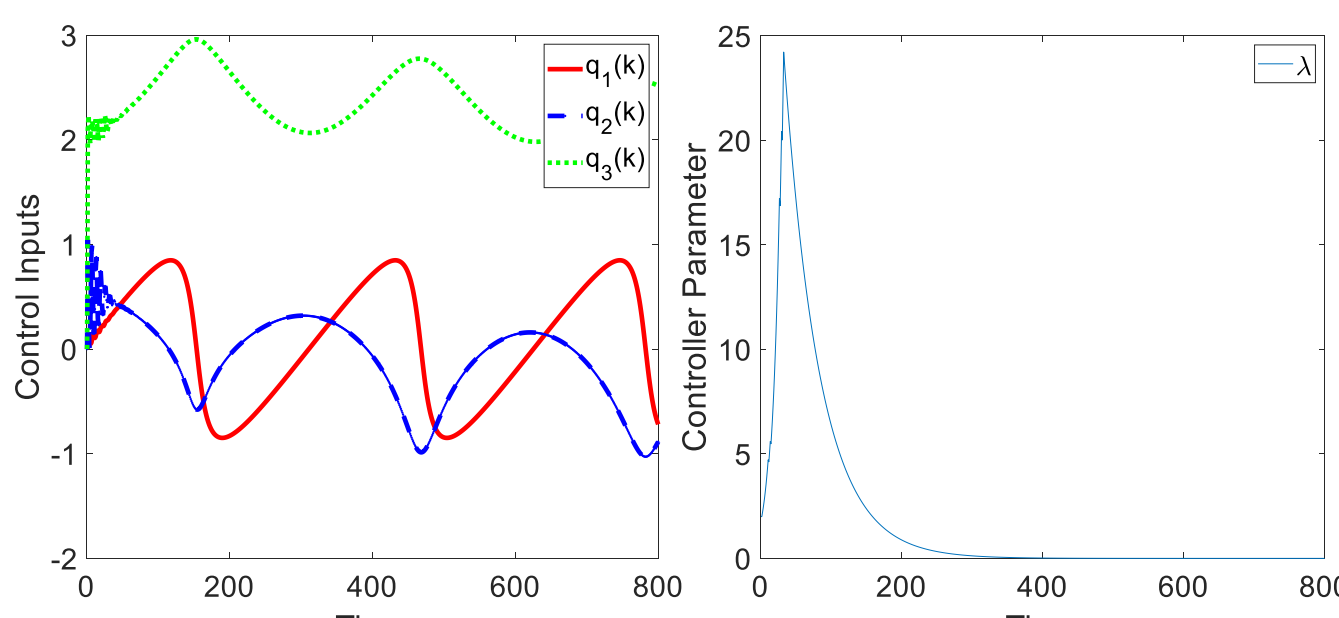

Fig. 3 Control inputs and controller parameter λ

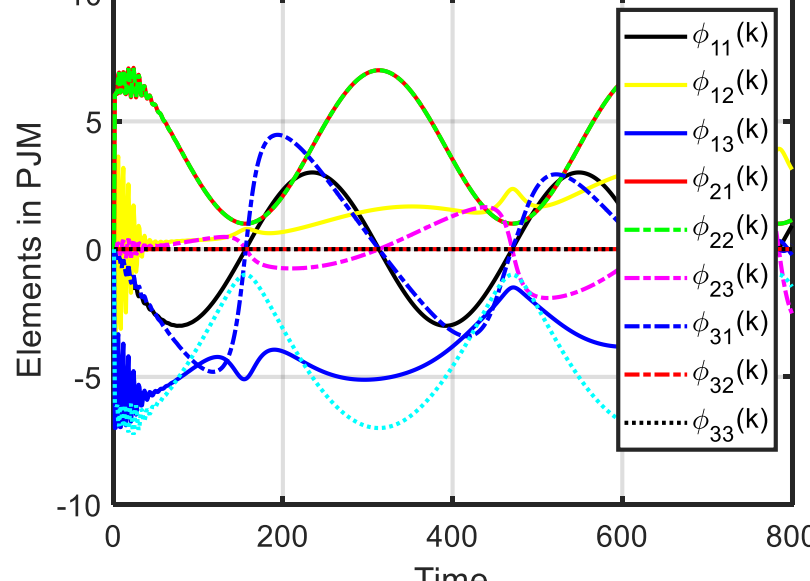

Fig. 4 Elements in $\boldsymbol{J}(k)$

Example 2: A six dimensional industrial robot is shown in Fig. 5.

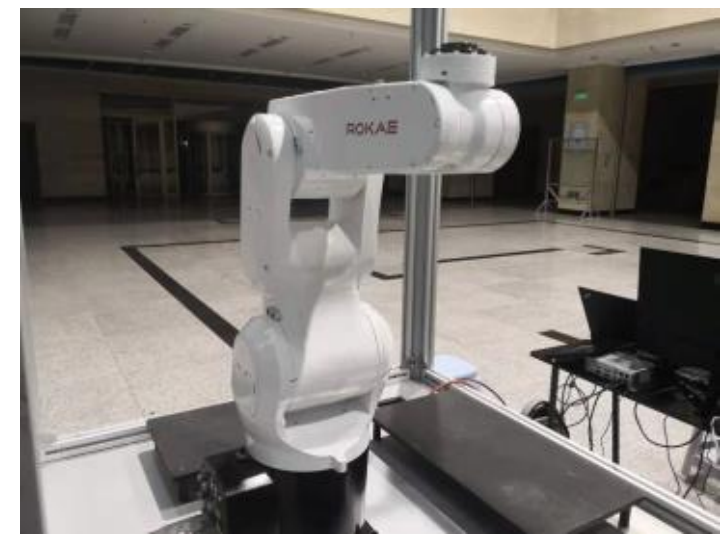

Fig. 5 Industrial robot

Herein, we will design the MFAPC controller for the industrial robot manipulator. The Denavit-Hartenberg parameters of the manipulator are listed in [14] to obtain the frame in Cartesian-space.

**Part A: Iterative MFAPC controller design**

The path generator is designed for the linear spline with parabolic blends path according to [14]. The desired path (trajectory) point in Cartesian-space at path-planning time $k$ is

$$\boldsymbol{y}^*(k)=[x^*(k),y^*(k),z^*(k),\alpha^*(k),\beta^*(k),\gamma^*(k)]^T \tag{45}$$

Then the description of corresponding desired frame at path-planning time $k+j$ ($j$=1, ⋯, $n$) is written as

$$\boldsymbol{T}^*(k+j)=\begin{pmatrix}\boldsymbol{A}^*(k+j) & \boldsymbol{p}^*(k+j)\\ \boldsymbol{0}_{1\times 3} & 1\end{pmatrix} \tag{46}$$

where

$$\boldsymbol{A}^*(k+j)=\begin{bmatrix} t_{11}^*(k+j) & t_{12}^*(k+j) & t_{13}^*(k+j)\\ t_{21}^*(k+j) & t_{22}^*(k+j) & t_{23}^*(k+j)\\ t_{31}^*(k+j) & t_{32}^*(k+j) & t_{33}^*(k+j)\end{bmatrix} \tag{47}$$

denotes the desired rotation matrix and

$$\boldsymbol{p}^*(k+j)=[x^*(k+j),y^*(k+j),z^*(k+j)]^T \tag{48}$$

represents the desired position vector at path-planning time $k+j$. According to [17] and [14], we can easily calculate $t_{11}^*(k+j)$, ⋯, $t_{33}^*(k+j)$ in accordance with (45) at time $k+j$.

Assume that the calculated robot frame at the $i$-th iteration for the inverse kinematics of the desired path vector $\boldsymbol{Y}_n^*(k+1)$ is described by

$$\boldsymbol{T}(k)_{(i)}=\begin{bmatrix} t(k)_{11(i)} & t(k)_{12(i)} & t(k)_{13(i)} & x(k)_{(i)}\\ t(k)_{21(i)} & t(k)_{22(i)} & t(k)_{23(i)} & y(k)_{(i)}\\ t(k)_{31(i)} & t(k)_{32(i)} & t(k)_{33(i)} & z(k)_{(i)}\\ 0 & 0 & 0 & 1\end{bmatrix} \tag{49}$$

The corresponding orientation matrix is written as

$$\boldsymbol{A}(k)_{(i)}=\begin{bmatrix} t(k)_{11(i)} & t(k)_{12(i)} & t(k)_{13(i)}\\ t(k)_{21(i)} & t(k)_{22(i)} & t(k)_{23(i)}\\ t(k)_{31(i)} & t(k)_{32(i)} & t(k)_{33(i)}\end{bmatrix} \tag{50}$$

The desired orientation matrix at path-planning time $k+j$ relative to the calculated orientation matrix of manipulator at the $i$-th iteration will be

$$\boldsymbol{D}(k+j)_{(i)}=\begin{bmatrix} d(k+j)_{11(i)} & d(k+j)_{12(i)} & d(k+j)_{13(i)}\\ d(k+j)_{21(i)} & d(k+j)_{22(i)} & d(k+j)_{23(i)}\\ d(k+j)_{31(i)} & d(k+j)_{32(i)} & d(k+j)_{33(i)}\end{bmatrix} =\boldsymbol{A}^*(k+j)\left(\boldsymbol{A}(k)_{(i)}\right)^{-1} \tag{51}$$

Here, the orientation matrix is converted into equivalent angle-axis representation to calculate the difference among the Euler angles $\alpha^*(k+j)-\alpha(k)_{(i)}$, $\beta^*(k+j)-\beta(k)_{(i)}$ and $\gamma^*(k+j)-\gamma(k)_{(i)}$. The equivalent angle is

$$\theta(k+j)_{(i)}=\arccos\left(\frac{d(k+j)_{11(i)}+d(k+j)_{22(i)}+d(k+j)_{33(i)}-1}{2}\right) \tag{52}$$

The equivalent axis of a finite rotation is

$$\hat{\boldsymbol{K}}(k+j)_{(i)}=\begin{pmatrix} d(k+j)_{32(i)}-d(k+j)_{23(i)}\\ d(k+j)_{13(i)}-d(k+j)_{31(i)}\\ d(k+j)_{21(i)}-d(k+j)_{12(i)}\end{pmatrix}\Bigg/ 2\sin\theta(k+j)_{(i)} \tag{53}$$

Then the relative rotational Euler angles vector is calculated by

$$\begin{bmatrix}\alpha^*(k+j)-\alpha(k)_{(i)}\\ \beta^*(k+j)-\beta(k)_{(i)}\\ \gamma^*(k+j)-\gamma(k)_{(i)}\end{bmatrix}=\hat{\boldsymbol{K}}(k+j)_{(i)}\theta(k+j)_{(i)} \tag{54}$$

The control law is

$$\begin{aligned}\Delta\boldsymbol{Q}(k)_{(i)}&=[\Delta\boldsymbol{q}^T(k)_{(i)},\cdots,\Delta\boldsymbol{q}^T(k+n-1)_{(i)}]^T\\ &=[\bar{\boldsymbol{\Psi}}^T(k)_{(i)}\bar{\boldsymbol{\Psi}}(k)_{(i)}+\bar{\boldsymbol{\lambda}}(k)_{(i)}]^{-1}\bar{\boldsymbol{\Psi}}^T(k)_{(i)}[\boldsymbol{Y}_n^*(k+1)-\boldsymbol{E}_{\substack{(n\bullet My)\\ \times My}}\boldsymbol{y}(k)_{(i)}]\\ &=[\bar{\boldsymbol{\Psi}}^T(k)_{(i)}\bar{\boldsymbol{\Psi}}(k)_{(i)}+\bar{\boldsymbol{\lambda}}(k)_{(i)}]^{-1}\bar{\boldsymbol{\Psi}}^T(k)_{(i)}[x^*(k+1)-x(k)_{(i)},\\ &\quad y^*(k+1)-y(k)_{(i)},z^*(k+1)-z(k)_{(i)},\alpha^*(k+1)-\alpha(k)_{(i)},\\ &\quad \beta^*(k+1)-\beta(k)_{(i)},\gamma^*(k+1)-\gamma(k)_{(i)},\cdots,x^*(k+n)-x(k)_{(i)},\\ &\quad y^*(k+n)-y(k)_{(i)},z^*(k+n)-z(k)_{(i)},\alpha^*(k+n)-\alpha(k)_{(i)},\\ &\quad \beta^*(k+n)-\beta(k)_{(i)},\gamma^*(k+n)-\gamma(k)_{(i)}]^T\end{aligned} \tag{55}$$

where

$$\begin{bmatrix}\boldsymbol{q}(k)_{(i)}\\ \vdots\\ \boldsymbol{q}(k+n-1)_{(i)}\end{bmatrix}=\begin{bmatrix}\boldsymbol{q}(k)_{(i-1)}\\ \vdots\\ \boldsymbol{q}(k)_{(i-1)}\end{bmatrix}+\begin{bmatrix}\boldsymbol{I} & & \\ \vdots & \ddots & \\ \boldsymbol{I} & \cdots & \boldsymbol{I}\end{bmatrix}\Delta\boldsymbol{Q}(k)_{(i)} \tag{56}$$

and we can easily obtain the Jacobian matrix $\boldsymbol{J}(k+j)_{(i+1)}$, ($j$=1, ⋯,$n$-1) in accordance with (35) and (56) to constitute $\bar{\boldsymbol{\Psi}}^T(k)_{(i+1)}$.

Then we choose the local optimal solution for the controller

$$\boldsymbol{q}(k)_{(i)}=\boldsymbol{q}(k)_{(i\text{-}1)}+\boldsymbol{g}^T\Delta\boldsymbol{Q}(k)_{(i)} \tag{57}$$

According to the robotic kinematics (34), we can easily transform (56) to

$$\boldsymbol{Y}_n(k+1)_{(i+1)}\triangleq[\boldsymbol{y}^T(k)_{(i+1)},\cdots,\boldsymbol{y}^T(k+n-1)_{(i+1)}]^T \tag{58}$$

and $\boldsymbol{y}(k)_{(i+1)}$ can be converted into (49) for the $i$+1-th iteration. Then (49)-(58) constitutes a closed-loop to find a solution for the minimum of $\left\|\boldsymbol{Y}_n^*(k+1)-\boldsymbol{Y}_n(k+1)_{(N)}\right\|_2$.

**Part B: Test MFAPC**

We have tried to design MFAPC with prediction step $n$=3. Since the calculation time of $[\bar{\boldsymbol{\Psi}}^T(k)_{(i)}\bar{\boldsymbol{\Psi}}(k)_{(i)}+\bar{\boldsymbol{\lambda}}(k)_{(i)}]_{18\times 18}^{-1}$ exceeds 1ms which is the implementation period of the robot, the industrial computer does not send any signals to the robot.

To reduce the amount of calculation of inverse operation, we design the iterative MFAPC with $n$=2 and $\bar{\boldsymbol{\lambda}} = \mathbf{0}$ as follow

$$\begin{aligned}\Delta \boldsymbol{Q}(k)_{(i)} &= [\Delta \boldsymbol{q}(k)_{(i)}, \Delta \boldsymbol{q}(k)_{(i+1)}] \\ &= \bar{\boldsymbol{\Psi}}^{-1}(k)_{(i)}[\boldsymbol{Y}_n^*(k+1) - \boldsymbol{E}_{\substack{(n\bullet My)\\ \times My}} \boldsymbol{y}(k)_{(i)}]\end{aligned} \tag{59}$$

where $\bar{\boldsymbol{\Psi}}^{-1}(k)_{(i)} = \begin{bmatrix} \boldsymbol{J}^{-1}(k)_{(i)} & \mathbf{0} \\ -\boldsymbol{J}^{-1}(k+1)_{(i)} & \boldsymbol{J}^{-1}(k+1)_{(i)} \end{bmatrix}$ for further simplifying the calculation of the inverse operation. And the maximum number of iterations is set to 10.

The beginning joint angle vector of manipulator is $A[-\pi/4, \pi/6, 0, 0, -\pi/4, 0]^T$ and the goal frame is set to $\boldsymbol{B}$ in Cartesian space whose corresponding joint angle vector is $[\pi/4, \pi/6, 0, 0, -\pi/4, 0]^T$ .

Fig. 6 shows the tracking performance of the manipulator and Fig. 7 shows the corresponding tracking error $\boldsymbol{y}^*(k+1) - \boldsymbol{y}(k)$ . Fig. 8 shows the measured joint angles and Fig. 9 shows the iteration count.

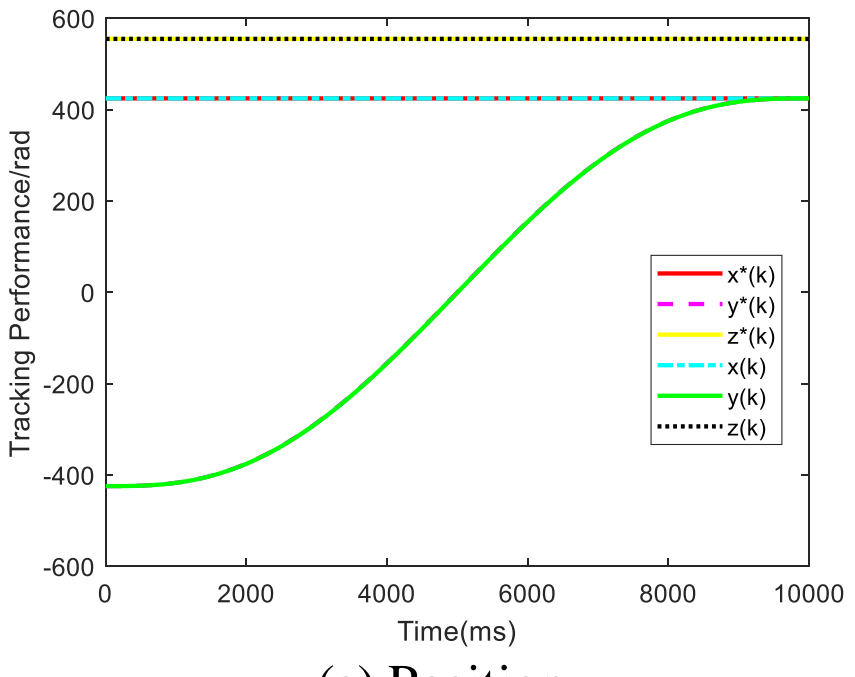

(a) Position

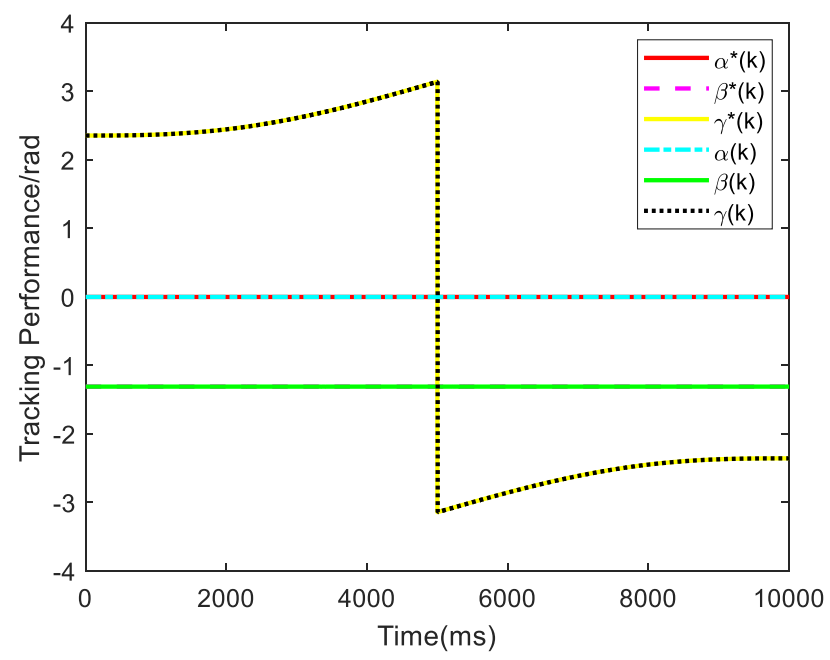

(b) Orientation

Fig. 6 Tracking performance

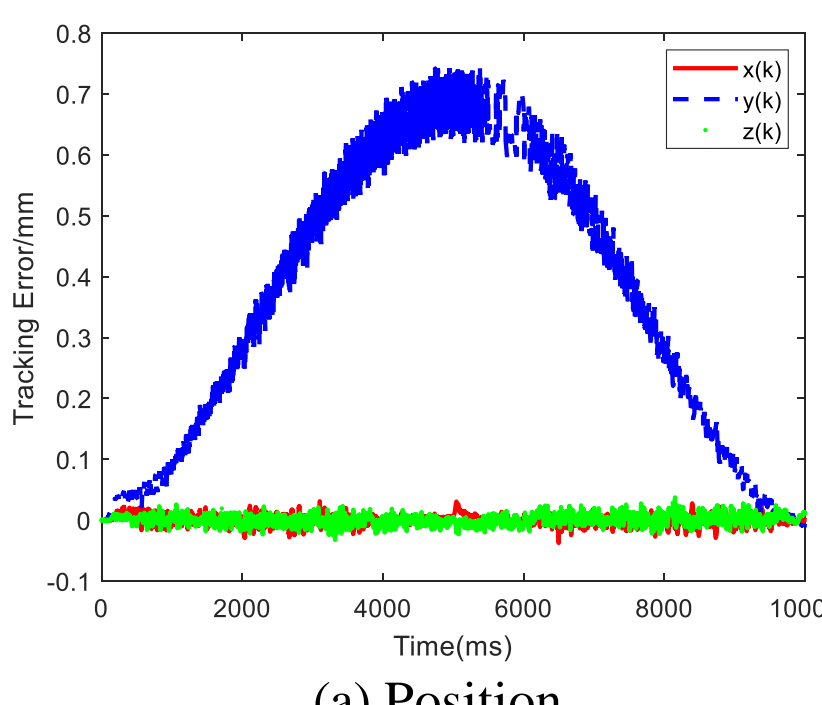

(a) Position

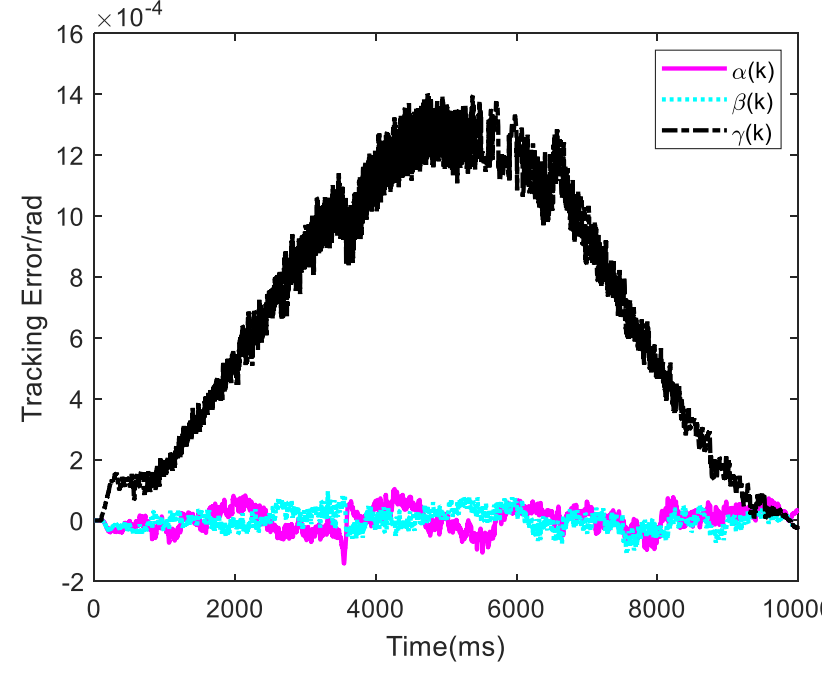

(b) Orientation

Fig. 7 Tracking error

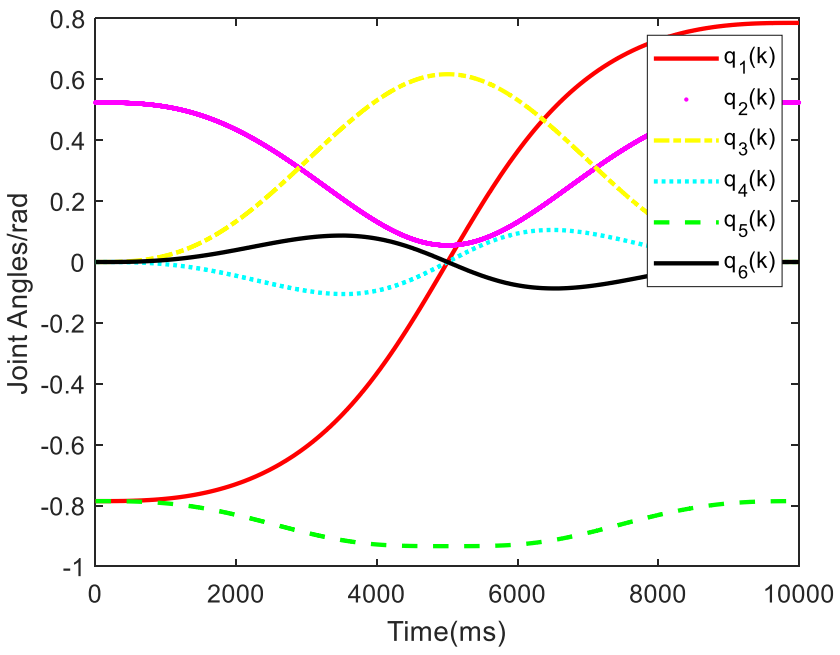

Fig. 8 Measured joint angles

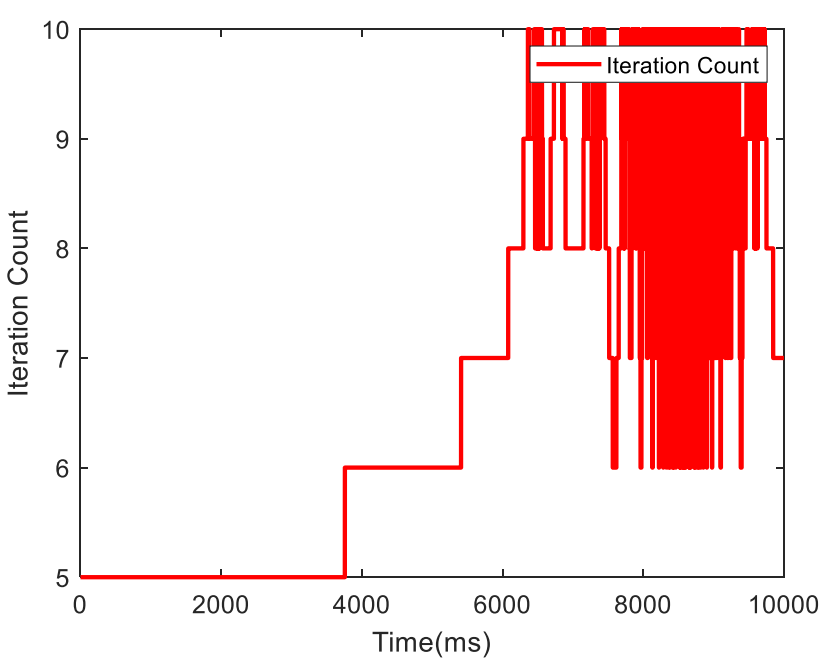

Fig. 9 Iteration count

## VI. CONCLUSION

In this brief, we have reanalyzed current robust numerical solution to the inverse kinematics based on Levenberg-Marquardt (LM) method through conventional control theory and reviewed its application in robotics. Compared to the current numerical analysis method, the relationship between the convergence performance of computational error and the damping factor is analyzed more clearly through analyzing the system control performance in this brief. Furthermore, by this effort, we also showed that the current works about MFAC are not studied in a reasonable way. Then we redesign the MFAC into MFAPC to utilize more desired trajectories in the future time. It not only shows an excellent control performance but also efficiently acquires the solution to the inverse kinematic of robotics.